\documentclass[aps,prb,floatfix,amsmath,amssymb,preprint,eqsecnum]{revtex4-1}
\usepackage{graphicx}
\usepackage{dcolumn}
\usepackage{bm}
\usepackage{epstopdf}
\usepackage{enumerate}
\usepackage{setspace}   
\usepackage{amsmath}
\usepackage{amssymb}

\begin{document}
\title{Effective theory of Fermi pockets in fluctuating antiferromagnets}

\author{Yang Qi}
\affiliation{Department of Physics, Harvard University, Cambridge MA
02138}

\author{Subir Sachdev}
\affiliation{Department of Physics, Harvard University, Cambridge MA
02138}

\date{\today\\
\vspace{1.6in}}
\begin{abstract}
We describe fluctuating two-dimensional 
metallic antiferromagnets by transforming to a rotating reference frame
in which the electron spin polarization is measured by its projections along the local antiferromagnetic order.
This leads to a gauge-theoretic description of an `algebraic charge liquid' involving spinless fermions and a spin $S=1/2$ complex scalar.
We propose a phenomenological effective lattice Hamiltonian which describes the binding of these particles into
gauge-neutral, electron-like excitations, and describe its implications for the electron spectral function across
the entire Brillouin zone. We discuss connections of our results to photoemission experiments
in the pseudogap regime of the cuprate superconductors.
\end{abstract}

\maketitle

\section{Introduction}
\label{sec:intro}

An understanding of the nature of the electron spectral function in the underdoped `pseudogap' regime
has emerged as one of the central problems in the study of the cuprate superconductors. 
A wealth of data has appeared in photoemission studies, some of which \cite{arc1,arc2,arc3,arc4,arc5,arc6} 
has been interpreted using
a model of ``Fermi arcs'' across the Brillouin zone diagonals; other studies \cite{pocket1,pocket2,pocket3}
have indicated the presence of pocket Fermi surfaces in the same region of the Brillouin zone.
Scanning tunnelling microscopy (STM) studies \cite{kohsaka1}  also indicate a Fermi arc of excitations which appears to end
abruptly at the magnetic Brillouin zone boundary associated with two sublattice N\'eel order.
Another issue of interest in experiments has been the angular dependence of the electronic excitation gap energy in the 
superconducting state. A `dichotomy' has been noted \cite{kohsaka1,dichot1,dichot2,dichot3} 
between the behavior of the gap near the nodal points on the
Brillouin zone diagonals, and the antinodal points along the principle axes of the Brillouin zone.

Related studies have also been made of the electronic spectra in the electron-doped 
cuprates \cite{armitage03,claesson04,matsui05,park07}. Here the spectra seem to be closer to 
that expected from conventional spin density wave theory, and so constitute an important testing
ground of theoretical ideas.

The recent observation of quantum oscillations in a strong magnetic 
field \cite{doiron,cooper,nigel,cyril,suchitra,louis,suchitra2} has given further impetus to
the development of a theory the normal state of the underdoped cuprates. Some of the transport data \cite{louis,chang,hackl1} has been
interpreted in terms of the presence of electron pockets in the hole-doped cuprates, for which there
is no apparent evidence in the photoemission studies; the latter, however, have only been carried out without a
magnetic field.

Finally, an important motivation for our study comes from numerical studies \cite{tremblay,dmft0,dmft1,dmft2,dmft3,dmft4} of Hubbard-like models for
the cuprate superconductors. These studies show a significant regime without any antiferromagnetic order, but with a
dichotomy in the normal state electronic spectra between the nodal and anti-nodal regimes. It would clearly be useful to have
analytic effective models which can be used to interpret the numerical data, and we shall propose such models here.

The theory we present here builds upon the framework set up in Refs.~\onlinecite{rkk1,rkk2}. These papers 
described a non-Fermi liquid state which was labeled an `algebraic charge liquid' (ACL); this state was obtained
after the loss of antiferromagnetic order in a metal via an unconventional transition \cite{senthil1} . 
As we will review below, in its simplest realization, the degrees of freedom of the ACL are spinless fermions and 
a $S=1/2$ complex boson $z_\alpha$ which interact via an emergent U(1) gauge force. This gauge force
has strong effects even in possible deconfined phases, and can lead to the formation of electron-like
bound states between the fermions and $z_\alpha$. The main purpose of this paper is to present
a general discussion of the dynamics of these fermionic bound states, and their influence on the photoemission spectra.
A full description of the pseudogap regime in the hole-doped cuprates 
will also require \cite{gs} considerations of pairing of these fermionic bound
states into bosonic Cooper pairs: this we defer to a subsequent paper. 

In our previous work \cite{rkk1,rkk2}, we presented analytic
arguments based upon the structure of a continuum theory valid at long wavelengths. Here, we shall present a more general
formulation of the theory, which allows computation of the electronic spectrum across the entire Brillouin zone. We shall show how
arguments based upon symmetry and gauge invariance allow construction of an effective theory for the electronic spectrum.
The theory will contain a number of coupling constants, whose values will have to be determined by comparing to numerical studies or experiments. Also, while the previous work \cite{rkk1,rkk2} used strong-coupling perspective, starting from the Schwinger boson theory of the antiferromagnet.
It is possible to derive the results presented below also from this strong-coupling approach. However, we will choose to present
our results by departing \cite{su2} from the ``spin-fermion'' model, which was originally developed from the weak coupling expansion.

Let us begin by defining the Lagrangian, $\mathcal{L}_{sf}$, of the spin-fermion model. \cite{spinfermion} We consider fermions $c_{i \alpha}$ $(\alpha, \beta = \uparrow, \downarrow )$
hopping on the sites of a square lattice. These are coupled to the fluctuations of the unit vector field $n^a_i$ 
($a=x,y,z$) representing the local orientation of the collinear antiferromagnetic N\'eel order. We will restrict our attention here to antiferromagnetic
order at wavevector ${\bf K} = (\pi,\pi)$, although generalizations to other ${\bf K}$ are possible. Throughout, we will freely make a gradient
expansion of $n^a_i$ over spatial co-ordinates $r = (x_i, y_i)$, focusing on the long wavelength fluctuations of the order parameter.
However, the fermion fields $c_{i, \alpha}$ have important low energy modes at many locations in the Brillouin zone, and so we
will not make any gradient expansion on the fermion operators.
We have the imaginary time ($\tau$)
Lagrangian
\begin{eqnarray}
\mathcal{L}_{sf} &=& \mathcal{L}_c + \mathcal{L}_\lambda + \mathcal{L}_n \nonumber \\
\mathcal{L}_c &=& \sum_i c_{i \alpha}^\dagger (\partial_\tau - \mu) c_{i \alpha} - \sum_{i<j} t_{ij} \left( c^{\dagger}_{i \alpha} c_{j \alpha} + c^{\dagger}_{j \alpha} c_{i \alpha} \right) \nonumber \\
\mathcal{L}_\lambda &=& - \lambda \sum_i (-1)^{x_i + y_i} n_i^a c^\dagger_{i \alpha} \sigma^a_{\alpha\beta} c_{i \beta} \nonumber \\ 
\mathcal{L}_n &=& 
\frac{1}{2g} \int d^2 r \left[ (\partial_\tau n^a )^2 + v^2 (\partial_r n^a)^2 \right].
\label{esf}
\end{eqnarray}
Here $t_{ij}$ are arbitrary hopping matrix elements describing the ``large'' Fermi surface, $\mu$ is the chemical potential, $\lambda$
is the spin-fermion coupling, $g$ controls the strength of the antiferomagnetic fluctuations,
$\sigma^a$ are the Pauli
matrices, and the $n^a$ field obeys the local constraint $\sum_a (n^a)^2 = 1$. Almost all previous
studies of the spin-fermion model \cite{spinfermion} have involved a perturbative treatment in powers of the coupling
$\lambda$, along with resummations of this expansion. Here, we will not expand in powers of $\lambda$, treating it
is a coupling of order unity: instead our analysis are motivated by expansions either in the number of field components,
or by small $g$.

For sufficiently small values of the coupling $g$, the model $\mathcal{L}_{sf}$ clearly has an antiferromagnetically ordered spin density wave (SDW) ground state with $\langle n^a \rangle \neq 0$. We are interested here in the mechanism by which
this order is lost as $g$ is increased, 
and a metallic state with no broken symmetries is obtained. In a recent paper \cite{su2} with others, we 
argued that there were 2 generic possibilities:\\
({\em i\/}) In the first case, there was a direct transition at a single critical $g=g_c$ to 
a Fermi liquid metal with a large Fermi surface. 
This transition has been examined in previous work \cite{spinfermion}, and is directly expressed in terms
of fluctuations of the O(3) order parameter $n^a$; the $T>0$ crossovers above $g=g_c$ have also been described \cite{tremblay,dmft0,dmft1,dmft2,dmft3}.\\ 
({\em ii\/}) The second possibility involved intermediate non-Fermi liquid phases before
the large Fermi surface metal was reached at sufficiently large $g$. In this case, the O(3) order parameter was parameterized in
terms of the spinor $z_\alpha$ by
\begin{equation}
n^a_i =  z_{i\alpha}^\ast \sigma^a_{\alpha\beta} z_{i \beta}. \label{nz}
\end{equation}
The spinor field $z_\alpha$ is the natural variable to describe the loss of magnetic order at 
$g=g_c$, and the non-Fermi liquid phases above $g_c$, and replaces the O(3) order parameter $n^a$.

Our focus in the present paper will be on the second possibility. Part of our motivation comes from
transport measurements \cite{hussey,louislinear}, which show an extended regime of non-Fermi liquid
behavior as $T \rightarrow 0$ in high magnetic fields.
We shall describe the photoemission spectra
at non-zero temperatures on both sides of $g_c$. Our $g \geq g_c$ results are candidates for the
pseudogap regime of the cuprates, and relate especially to recent experimental results of Meng {\em et al.} \cite{pocket3}.

To complement the approach taken in Ref.~\onlinecite{su2}, here we will motivate our choice of 
the non-magnetic non-Fermi liquid phases by extending the theory \cite{senthil1} of the loss of N\'eel order
in the insulator at half filling. Although
motivated in a theory of metals, $\mathcal{L}_{sf}$ also contains (in principle) 
a complete description of the insulating states at half-filling.
Crucial to the description of insulators,\cite{haldane,rs} are `hedgehog' point tunneling events (`instantons') in which $n^a$ points
radially outwards/inwards from a spacetime point. These hedgehogs carry Berry phases: in $\mathcal{L}_{sf}$ the Berry
phases are expected to appear from the determinant of the gapped fermions integrated out in a hedgehog field for $n^a$.
All previous treatments of the spin-fermion model have neglected the hedgehog Berry phases, and this may well
be appropriate under suitable circumstances in certain superconducting states.\cite{rkk3} 

In this paper, we wish to focus on regimes and phases where the hedgehog tunnelling events are suppressed. In the insulator,
the hedgehog suppression is a consequence of quantum interference from the hedgehog Berry phases, and leads to interesting
new `deconfined' phases and critical points.\cite{kamal,mv,senthil1} Hedgehog suppression is also possible in certain exotic
metallic states (to be described below), where fermionic excitations near a Fermi surface lead to a divergence in the
hedgehog action.\cite{hermele,sslee,rkk1,rkk2,rkk3}

\begin{figure}
  \includegraphics*[width=5in]{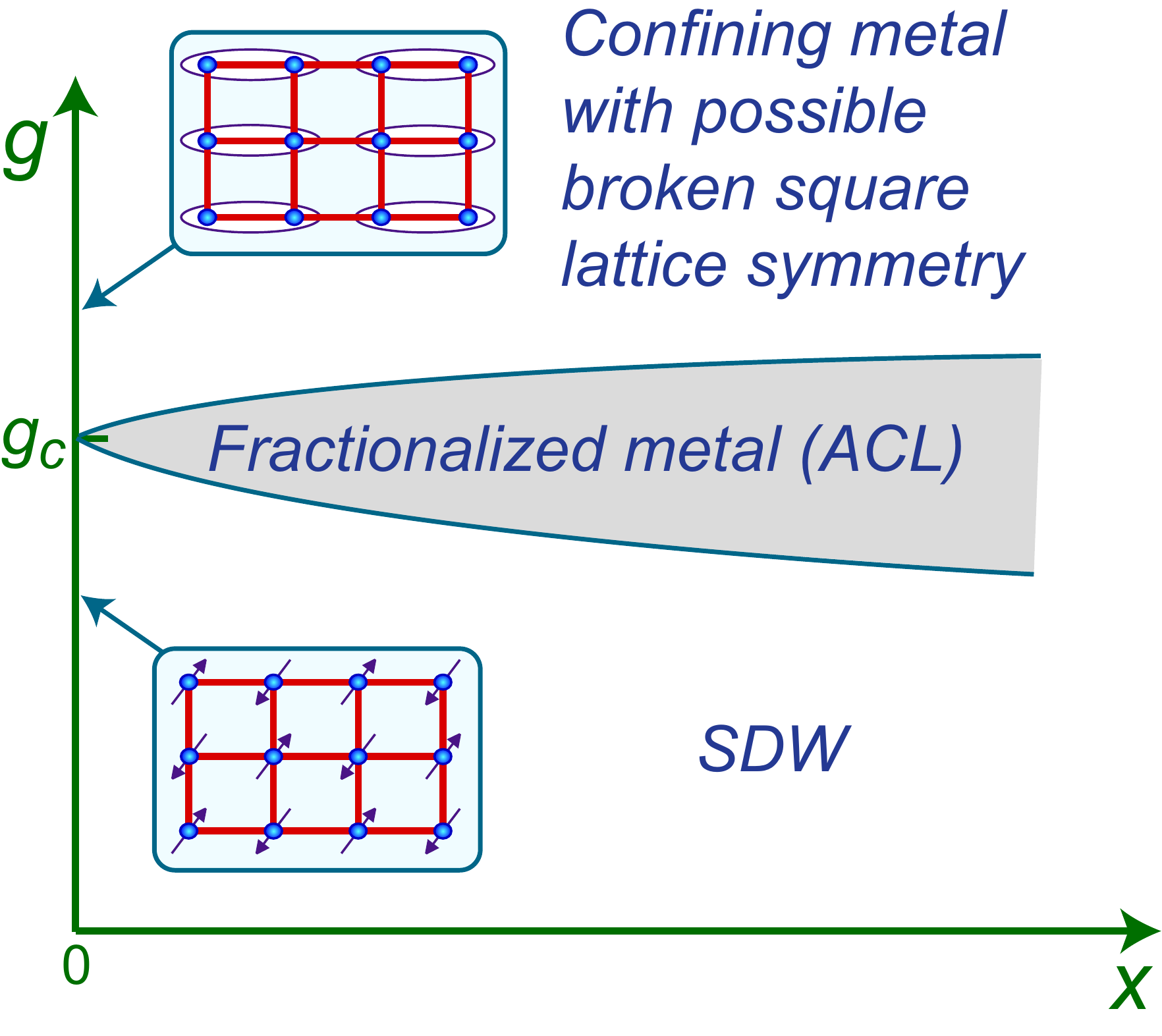}
  \caption{Proposed phase diagram \cite{rkk1} of a quantum antiferromagnet doped with carrier density $x$.
  The strength of antiferromagnetic fluctuations in the insulator is controlled by the coupling $g$.
  At $x=0$, there is a deconfined critical point \cite{senthil1}, separating a N\'eel state and a valence bond
  solid \cite{rs}. The broken symmetries in these two states survive in metallic Fermi liquid state at non-zero $x$.
  More importantly, the deconfined critical point broadens into a non-Fermi liquid phase, the algebraic charge liquid (ACL),
  which is the focus of attention in the present paper. The conventional large Fermi surface Fermi liquid is not
  shown in the phase diagram above. It appears at larger $x$, and its phase transitions to the ACL and the SDW states
  were discussed in earlier work \cite{su2}.}
  \label{figrkk}
\end{figure}

It is useful to begin our analysis by adapting the phase diagram of Ref.~\onlinecite{rkk1} describing the doping
of an insulating deconfined critical point -- see Fig.~\ref{figrkk}.
In the insulator, at $x=0$, the transition is between a N\'eel state and a valence bond solid.\cite{rs}
This transition is described \cite{senthil1} by a deconfined critical point, with the order parameter $n^a$ replaced by
a complex `relativistic' boson $z_\alpha$ which carries charges under an emergent U(1) gauge field $A_\mu$: this mapping will be reviewed
in Section~\ref{sec:acl}. Upon moving away from the insulator, the fermions $c_{\alpha}$ are replaced by fermions $\psi_p$
fermions which do not carry spin, but do carry charge $p=\pm1$ of the U(1) gauge field $A_\mu$. 
We will present the Lagrangian of $z_\alpha$, $\psi_p$, $A_\mu$ which described the phase diagram in Fig.~\ref{figrkk}
in Section~\ref{sec:acl}. 

In the SDW phase of Fig.~\ref{figrkk}, the $z_\alpha$, $\psi_p$, $A_\mu$ theory is formally identical
to the spin-fermion model of $c_\alpha$, $n^a$. This is because hedgehogs are strongly suppressed
when there is magnetic order. We explore this connection in Appendix~\ref{smallg}, and in Section~\ref{sec:rc}
to obtain the spectral functions of the electrons in the `renormalized classical' (RC) regime \cite{vilk,borejsza}.

The transition out of the SDW phase in Fig.~\ref{figrkk} is into a fractionalized metal phase, which was
dubbed\cite{rkk2} an algebraic charge liquid (ACL). This transition is in the O(4) universality class.\cite{rkk3}
To leading order, the elementary excitations of the ACL are simply the constituent fractionalized 
fields $z_\alpha$, $\psi_p$, $A_\mu$, and their interactions have been discussed in previous work.\cite{rkk3}

Here, our primary results will be on the fermionic spectrum over a large intermediate length scale
near the boundary between the ACL and the confining metallic states shown in Fig.~\ref{figrkk}.
A key characteristic of the deconfined critical point is parametrically large separation between the spin correlation
length, $\xi$, and the confinement scale, $\xi_{\rm conf}$, at which hedgehogs proliferate. In the doped system, as argued
in Ref.~\onlinecite{rkk1}, the fractionalized excitations are already bound into gauge neutral excitations
at the scale $\xi$. Among these bound states are electron-like fermions which carry charge $-e$ and spin $1/2$,
and so can be directly detected in photoemission experiments. We will present a phenomenological effective
Hamiltonian for these excitations in Section~\ref{sec:bound}.

\section{U(1) gauge theory}
\label{sec:acl}

The basic idea \cite{su2} of the mapping to the theory of the ACL is to
transform to a new set of fermions, $\psi_{ip}$ with $p =\pm 1$, with their spin components $p$
polarized along the direction of the local SDW order. We perform this rotation with the spacetime dependent SU(2) matrix
$R_{\alpha p}^i$ so that \cite{schulz,borejsza}
\begin{eqnarray}
c_{i \alpha} &=& R_{\alpha p}^i \psi_{i p}  \label{cab}
\end{eqnarray}
We choose $R_{\alpha p}$ so that spin-fermion coupling is only along $\sigma^z$, and so
\begin{equation}
n^a_i R^{i \dagger}_{p \alpha} \sigma^a_{\alpha\beta} R_{\beta p'}^i =  \sigma^z_{p p'} = p \delta_{p p'}
\end{equation}
This relationship is equivalent to 
\begin{equation}
n^a_i = \frac{1}{2} \mbox{Tr} \left( \sigma^a R^i \sigma^z R^{i\dagger} \right) \label{rz}
\end{equation}
Now, we parameterize
\begin{equation}
R^i = \left( \begin{array}{cc} z_{i\uparrow} & -z_{i\downarrow}^\ast \\
z_{i\downarrow} & z_{i\uparrow}^\ast \end{array} \right)
\end{equation}
with $\sum_\alpha |z_{i \alpha} |^2 = 1$.
We can verify that Eq.~(\ref{rz}) yields the usual relation in Eq.~(\ref{nz}).

A crucial feature of the resulting Hamiltonian for the $\psi_{i p}$ and $z_{i \alpha}$ is that it is
invariant under a local U(1) gauge transformation. This follows from the invariance of Eqs.~(\ref{cab}) and (\ref{nz})
under the transformation
\begin{eqnarray}
z_{i\alpha} &\rightarrow &z_{i\alpha} e^{i \vartheta_i} \nonumber \\
\psi_{i p} & \rightarrow & \psi_{i p} e^{- i p \vartheta_i}
\end{eqnarray}
where $\vartheta_i$ has an arbitrary dependence on space and time. Note that the $\psi_{i p}$ carry opposite charges
$p=\pm 1$ under the U(1) gauge transformation (which is unrelated to the gauge invariance associated with
the physical electromagnetic force). 
Associated with this U(1) gauge invariance, we will introduce an internal dynamical gauge field $A_\mu$
in constructing the effective theory. 

Ref.~\onlinecite{su2} argued that describing the transition to the
large Fermi surface Fermi liquid required the inclusion of additional degrees of freedom so that the theory
had a SU(2) gauge invariance. We will not consider this extension here. However, we expect that simple
extensions of the results in Section~\ref{sec:bound} apply also to the SU(2) ACL phases found in Ref.~\onlinecite{su2}.

We can now insert Eqs.~(\ref{cab}) and (\ref{nz}) into Eqs.~(\ref{esf}) and obtain the
theory of fluctuating Fermi pockets. We will assume that the $z_{i \alpha}$ are slowly varying, but allow the 
fermion fields $\psi_{i p}$ to have an arbitrary dependence on spacetime. The complete Lagrangian is written as
\begin{equation}
\mathcal{L}_{acl} = \mathcal{L}_z + \mathcal{L}_{\psi} + \mathcal{L}_{ss} \label{lacl}
\end{equation}
The first term is the CP$^1$ model for the $z_\alpha$:
\begin{equation}
\mathcal{L}_z = \frac{2}{g} \left[ | (\partial_\tau - i A_\tau) z_\alpha |^2 + v^2 |({\bm \nabla} - i {\bf A} ) z_\alpha|^2 \right]
\end{equation}
The fermion hopping term in Eq.~(\ref{esf}) yields some interesting structure; it can be written as
\begin{eqnarray}
&& - \sum_{i<j} t_{ij} \Biggl[   \bigl( z_{i \alpha}^\ast z_{j \alpha} \bigr) \left( \psi^\dagger_{i+} \psi_{j+} + \psi^\dagger_{j-} \psi_{i-} \right)
\nonumber \\
&&~~~~~~~~~+ \bigl( z_{j \alpha}^\ast z_{i \alpha} \bigr) \left( \psi^\dagger_{i-} \psi_{j-} + \psi^\dagger_{j+} \psi_{i+} \right) \nonumber \\
&&~~~~~~~~~+\bigl( \varepsilon^{\alpha\beta} z_{j \alpha}^\ast z_{i \beta}^\ast \bigr) \left( \psi^\dagger_{i+} \psi_{j-}
- \psi_{j+}^\dagger \psi_{i-} \right) \nonumber \\
&&~~~~~~~~~+ \bigl( \varepsilon^{\alpha\beta} z_{i \alpha} z_{j \beta} \bigr) \left( \psi^\dagger_{i-} \psi_{j+} 
- \psi^\dagger_{j-} \psi_{i+} \right)
\Biggr] \label{hop}
\end{eqnarray}
Now, from the derivation of the CP$^1$ model we know that
\begin{equation}
z_{i \alpha}^\ast z_{j \alpha} \approx e^{i A_{ij}}
\end{equation}
and this is easily incorporated into the first two terms in Eq.~(\ref{hop}), yielding terms which are gauge invariant.
We therefore incorporate the first two terms in Eq.~(\ref{hop}) into the gauge-invariant Lagrangian
\begin{eqnarray}
\mathcal{L}_\psi &=& \sum_{p=\pm1} \sum_{i} \psi_{ip}^\dagger
\bigl( \partial_\tau + i p A_{\tau}  -\mu - \lambda p (-1)^{i_x+i_y} \bigr) \psi_{ip}  \nonumber \\
&~&- \sum_{p=\pm 1} \sum_{i<j} t_{ij} \left( e^{i p A_{ij}} \psi_{ip}^\dagger \psi_{jp} + e^{-i p A_{ij}} \psi_{jp}^\dagger \psi_{ip} \right)
\label{lpsi}
\end{eqnarray}
For $A_\mu = 0$, $\mathcal{L}_\psi$ describes the band structure in terms of the Fermi pockets. 
The interactions arise from the minimal coupling to the $A_\mu$ gauge field.
Finally, we need to consider the last two terms in Eq.~(\ref{hop}). These are the analog of the `Shraiman-Siggia' couplings.\cite{shraiman}
Combining these terms with the analogous terms arising from the time derivative of the $c_\alpha$, we obtain
 to leading order
in the derivative of the $z_\alpha$:
\begin{eqnarray}
\mathcal{L}_{ss} &=&  \int_{{\bf k},{\bf p},{\bf q}} \left[ {\bf p} \cdot \frac{\partial \varepsilon({\bf k})}{\partial {\bf k}} \right] z_{\downarrow} ({\bf q}-{\bf p}/2) z_{\uparrow} ({\bf q}+{\bf p}/2) \psi^\dagger_- ({\bf k} + {\bf q}) \psi_+ ({\bf k} - {\bf q})  + \mbox{c.c.}
\nonumber \\
&~&~~~~~~~+ \sum_i (z_{i\uparrow} \partial_\tau z_{i\downarrow} - z_{i\downarrow} \partial_\tau z_{i\uparrow} ) \psi_{i-}^\dagger
\psi_{i+} + \mbox{c.c.}
\label{lss}
\end{eqnarray}
where $\varepsilon({\bf k})$ is the single particle dispersion of the large Fermi surface state:
\begin{equation}
\varepsilon ({\bf k} ) = - \sum_j t_{ij} e^{i {\bf k} \cdot ({\bf r}_j - {\bf r}_i)}.
\end{equation}
Note that the terms in $\mathcal{L}_{ss}$ mix fermions with different $A_\mu$ charges.

\begin{table}[t]
\begin{spacing}{2}
\centering
\begin{tabular}{||c||c|c|c|c||} \hline\hline
 & $T_x$ & $R_{\pi/2}^{\rm dual}$ & $I_x^{\rm dual}$ & $\mathcal{T}$  \\
 \hline\hline
$~~z_\alpha$~~ & $ ~\epsilon_{\alpha\beta} z^{\beta \ast}~$
& $ ~\epsilon_{\alpha\beta} z^{\beta \ast}~$ & $
~\epsilon_{\alpha\beta} z^{\beta \ast}~$ & $
~\epsilon_{\alpha\beta} z^{\beta \ast}~$
\\
\hline
$\psi_{+}$ & $ -\psi_{-}$ & $ -\psi_{-}$ & $ -\psi_{-}$  &  $\psi_{+}^\dagger $ \\
\hline $\psi_{-}$ & $\psi_{+}$ & $ \psi_{+}$ & $ \psi_{+}$  & $
\psi_{-}^\dagger$ 
\\
\hline
$A_\tau$ & $-A_\tau$ & $-A_\tau$ & $-A_\tau$ & $A_\tau$   
\\
\hline
$A_x$ & $-A_x$ & $-A_y$ & $A_x$ & $-A_x$   
\\
\hline
$A_y$ & $-A_y$ & $A_x$ & $-A_y$ & $-A_y$   
\\
\hline
$n^a$ & $-n^a$ &  $-n^a$ &  $-n^a$ &  $-n^a$   
\\
\hline
$F_\alpha$ & $G_\alpha$ &  $G_\alpha$ &  $G_\alpha$ &  $\epsilon^{\alpha\beta}F_\beta^\dagger$   
\\
\hline
$G_\alpha$ & $F_\alpha$ &  $F_\alpha$ &  $F_\alpha$ &  $\epsilon^{\alpha\beta}G_\beta^\dagger$   
\\ \hline\hline
\end{tabular}
\end{spacing}
\caption{Transformations of the lattice fields under square
lattice symmetry operations. $T_x$: translation by one lattice
spacing along the $x$ direction; $R_{\pi/2}^{\rm dual}$: 90$^\circ$
rotation about a dual lattice site on the plaquette center
($x\rightarrow y,y\rightarrow-x$); $I_x^{\rm dual}$: reflection
about the dual lattice $y$ axis ($x\rightarrow -x,y\rightarrow y$);
$\mathcal{T}$: time-reversal, defined as a symmetry (similar to
parity) of the imaginary time path integral. Note that such a
$\mathcal{T}$ operation is not anti-linear. The transformations of
the Hermitian conjugates are the conjugates of the above, except for
time-reversal of fermions \cite{vortexpsg}. For the latter,
$\psi_{\pm}$ and $\psi^\dagger_{\pm}$ are treated as independent Grassman
numbers and $\mathcal{T}: \psi^\dagger_{\pm} \rightarrow - \psi_{\pm}$; similarly for
$F_\alpha$, $G_\alpha$.}
\label{table0}
\end{table}
The analysis in the following Section~\ref{sec:bound} will be based largely on symmetry,
and so it is useful to recall \cite{rkk1,vortexpsg} now how the fields introduced so far transform under symmetry operations.
These are summarized in Table~\ref{table0}.

\section{Effective theory of electrons and photons}
\label{sec:bound}

As discussed in Section~\ref{sec:intro}, 
 we want to work in the regime where the photon creates bound states between the 
$\psi$ fermions and the $z_\alpha$ spinons, but the monopole induced confinement has not yet occurred.
Thus we are in a fluctuating SDW state with a spin correlation length $\xi$, but we are interested in phenomena
at a scale larger than $\xi$. However, the confinement of the photons occurs at a scale $\xi_{\rm conf}$, and so we
will restrict ourselves to the $\xi < r < \xi_{\rm conf}$.

We also note the complementary considerations in the work of Wen and Lee \cite{leewen}: 
they considered a spin liquid model
with fermionic spinons and bosonic holons (in contrast to our bosonic spinons and spinless fermion charge carriers),
and described spinon-holon bound states in the electron spectral
functions. Also, Essler and Tsvelik \cite{essler02,essler05} described a model of weakly coupled
chains, and considered the bound states of the spinons and holons of the one-dimensional spin liquid on the chains.
In our approach, we do not appeal to these spin liquid states,
but deal instead with states motivated by the fluctuations in the observed spin density wave order.

The regime $\xi < r < \xi_{\rm conf}$
was treated in Section IV of Ref.~\onlinecite{rkk1}. See also the subsection on holon-spinon binding
in the Appendix of Ref.~\onlinecite{rkk2}. Here we shall provide a more general treatment, which should also
allow for a computation of spectral functions.

Let the bound state between the $\psi_+$ fermions and the $z_\alpha$ be $F_\alpha$.
The bound state should have the full symmetry of the square lattice, and so we can define
a local operator $F_{i \alpha}$, which creates this bound state centered at the lattice site $i$.

As was emphasized in the initial analysis\cite{rkk1}, there is a second independent bound state,
and a consequent doubling of the fermion species. This is the bound state between the $\psi_-$ and
$z_\alpha^\ast$, which we denote by the local operator $G_{i \alpha}$. In the ordered N\'eel state, the sublattice
location of a fermion also fixes its spin. However, when we move to length scales larger than $\xi$, this is no longer
the case because the spin direction of the background N\'eel state has been averaged over.
Thus we can view $F_\alpha$ and $G_\alpha$ as fermions that reside preferentially (but not exclusively) on the
two sublattices, and they separately have an additional degeneracy associated with carrying $S=1/2$.

More formally, all we will really need are the properties of $F_{\alpha}$ and $G_{\alpha}$ under
the square lattice symmetry operations: these are summarized in Table~\ref{table0}, and will form the 
bases of our analysis below.

The bare electron operator will have a non-zero overlap with both the $F_\alpha$ and $G_\alpha$ fermions.
This will be non-local over the scale $\xi$. We approximate this connection from Eq.~(\ref{cab}) 
as
\begin{eqnarray}
c_{i \alpha} &=& z_{i \alpha} \psi_{i +} - \epsilon_{\alpha\beta} z_\beta^\ast \psi_{i -} \nonumber \\
&\approx& Z \left( F_{i \alpha} + G_{i \alpha} \right) \label{cFG}
\end{eqnarray}
where $Z$ is some quasiparticle renormalization factor depending upon the fermion-spinon bound
state wavefunction. In general, $Z$ should be non-local over a scale $\xi$, but have limited ourselves for simplicity
to a momentum independent wavefunction renormalization. 
Note that Eq.~(\ref{cFG}) and the symmetry transformations in Table~\ref{table0} ensure that $c_{\alpha}$
is invariant under all operations of the square lattice symmetry. The possible non-local terms in Eq.~(\ref{cFG})
can be deduced by the requirements of symmetry.

We now need an effective Hamiltonian for the $F_{i \alpha}$ and $G_{i
  \alpha}$.  Formally, any Hamiltonian which is invariant under the
symmetry transformations of Table~\ref{table0} is acceptable; however,
we use simple physical requirements to restrict the large class of
possibilities.  For the diagonal terms which do not mix the $F$ and
$G$, we assume (for simplicity) that they just inherit the terms for
$\psi_+$ and $\psi_-$ in $\mathcal{L}_\psi$ in Eq.~(\ref{lpsi}). The
mixing between the $F$ and $G$ is provided by the Shraiman-Siggia term
$\mathcal{L}_{ss}$. Physically this can be understood as the mixing
corresponds to hopping between two sublattices (as $F$ and $G$ reside
preferentially on different sublattices), and Shraiman-Siggia term
describes such hopping and associated spin-flipping
process\cite{shraiman}.  These terms are more simply considered in
their real space form which are the last two terms in
Eq.~(\ref{hop}). Combining these terms, we have the effective
Hamiltonian
\begin{eqnarray}
H_{\rm eff} &=& - \sum_{ij} t_{ij} \left( F_{i \alpha}^\dagger F_{j \alpha} + G_{i \alpha}^\dagger G_{j \alpha} \right) 
- \lambda \sum_i (-1)^{i_x + i_y} \left( F_{i \alpha}^\dagger F_{i \alpha} - G_{i \alpha}^\dagger G_{i \alpha} \right) \nonumber \\
&~&~~~~~~- \sum_{ij} \widetilde{t}_{ij} \left( F_{i \alpha}^\dagger G_{j \alpha} + G_{i \alpha}^\dagger F_{j \alpha} \right) 
\label{heff}
\end{eqnarray}
where the second line comes from $\mathcal{L}_{ss}$. Here $t_{ij}$ and $\widetilde{t}_{ij}$ are renormalized in some
unknown manner from the bare hopping matrix elements in $\mathcal{L}_{sf}$. We can now verify that all the terms
above are invariant under the transformations in Table~\ref{table0}.

The above Hamiltonian of bound state can be diagonalized in momentum
space. First, rewrite equation (\ref{heff}) in momentum space
\begin{eqnarray}
   H_{\rm eff}&=&\sum_k(\varepsilon ({\bf k})-\mu)\left(F_{{\bf k} \alpha}^\dagger F_{{\bf k} \alpha}+
    G_{{\bf k} \alpha}^\dagger G_{{\bf k} \alpha}\right)
  +\sum_k\tilde{\varepsilon} ({\bf k})\left(F_{{\bf k} \alpha}^\dagger G_{{\bf k} \alpha}+
    G_{{\bf k} \alpha}^\dagger F_{{\bf k} \alpha}\right) \nonumber \\
    &~&~~~~~~~~~~~~~
  -\lambda\left(F_{{\bf k} \alpha}^\dagger F_{{\bf k} + {\bf K}, \alpha}-
    G_{{\bf k} \alpha}^\dagger G_{{\bf k} + {\bf K}, \alpha}\right)
 \label{eq:H-bs}
\end{eqnarray}
where we parameterize $\varepsilon ({\bf k})$ and $\tilde{\varepsilon} ({\bf k})$ as
\begin{align}
  \label{eq:ek}
  \varepsilon ({\bf k})&=-2t(\cos k_x+\cos k_y)+4t^\prime\cos k_x\cos k_y
  -2t^{\prime\prime}(\cos2k_x+\cos2k_y)\\
  \label{eq:tek}
  \tilde{\varepsilon} ({\bf k})&=-\tilde{t}_0-2\tilde{t}(\cos k_x+\cos k_y)+
  4\tilde{t}^\prime\cos k_x\cos k_y
  -2\tilde{t}^{\prime\prime}(\cos2k_x+\cos2k_y)
\end{align}
Here $t$, $t^\prime$ and $t^{\prime\prime}$ are nearest neighbor, next
nearest neighbor and next-next-nearest neighbor hopping $t_{ij}$
respectively. $\tilde{t}$, $\tilde{t}^\prime$ and
$\tilde{t}^{\prime\prime}$ are the hopping elements of
$\tilde{t}_{ij}$. Here we only included terms up to third nearest
neighbor hopping, which is capable to capture the shape of the Fermi
surface, but higher order terms can be included in a similar
fashion. $\tilde{t}_0$ is the matrix element of the on-site mixing
term $F_i^\dagger G_i+G_i^\dagger F_i$, which is also allowed by symmetry.

To diagonalize Hamiltonian (\ref{eq:H-bs}), we change basis to
\begin{equation}
  \label{eq:CD}
  C_{{\bf k} \alpha}=\frac{1}{\sqrt{2}}(F_{{\bf k} \alpha}+G_{{\bf k} \alpha}),
  \quad D_{{\bf k} \alpha}=\frac{1}{\sqrt{2}}(F_{{\bf k} + {\bf K}, \alpha}-G_{{\bf k} + {\bf K}, \alpha})
\end{equation}
Note that $D_{{\bf k}\alpha}$ has momentum ${\bf k}$ according to the transformation under
translation symmetry listed in Table \ref{table0}. In the new basis,
the Hamiltonian becomes
\begin{eqnarray}
  H_{\rm eff} &=&\sum_k\left[(\varepsilon ({\bf k})+\tilde{\varepsilon} ({\bf k})-\mu)C_{{\bf k} \alpha}^\dagger C_{{\bf k} \alpha}
  +(\varepsilon ({\bf k} + {\bf K}) -\tilde\varepsilon ({\bf k} + {\bf K}) -\mu)D_{{\bf k} \alpha}^\dagger D_{{\bf k} \alpha} \right.
  \nonumber \\
  &~&~~~~\left. -\lambda\left(C_{{\bf k} \alpha}^\dagger D_{{\bf k} \alpha}+D_{{\bf k} \alpha}^\dagger C_{{\bf k} \alpha}\right)\right]
  \label{eq:H-bs-cd}
\end{eqnarray}

The spectrum of electron operator $c$ can be obtained by diagonalizing the
Hamiltonian of $C$ and $D$ fermions, as $c$ is related to $C$ operator
according to equation (\ref{cFG}) $c_{k\alpha}\simeq
({Z}/{\sqrt{2}}) C_{{\bf k} \alpha}$. In the Hamiltonian (\ref{eq:H-bs-cd}),
$C$ and $D$ fermions have dispersions $\varepsilon ({\bf k})+\tilde{\varepsilon} ({\bf k})$
and $\varepsilon ({\bf k} + {\bf K}) -\tilde\varepsilon ({\bf k} + {\bf K}) $ respectively, and they are
mixed through the $\lambda$ term.  With the mixing, gaps will open
where the Fermi surfaces of $C$ and $D$ fermions intersect and the
large Fermi surfaces become Fermi pockets. In the case of
$\tilde{\varepsilon} ({\bf k})=0$, the $D$ Fermi surface is the same as the $C$
Fermi surface shifted by $(\pi,\pi)$; so the pockets are symmetric
under reflection with respect to the magnetic Brillouin zone boundary,
and therefore centered at $(\pi/2, \pi/2)$. However, with a
non-vanishing $\tilde{\varepsilon} ({\bf k})$, the dispersions of $C$ and $D$ are
different, so the pockets are no longer symmetric about the magnetic
Brillouin zone boundary, and are not necessarily centered at $(\pi/2,
\pi/2)$.

\begin{figure}[htbp]
  \centering
  \includegraphics[width=\textwidth]{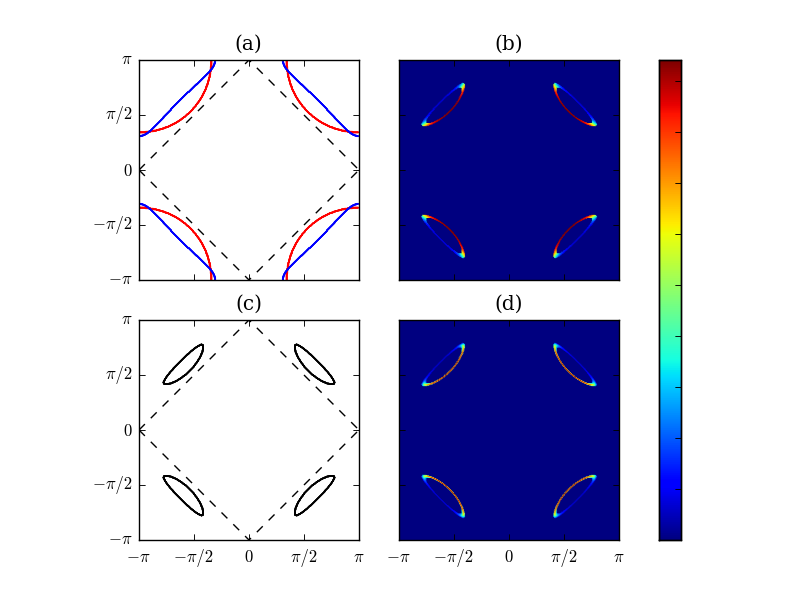}
  \caption{Plot of Fermi pockets of the bound state with
    $\tilde{\varepsilon} ({\bf k}) = 0.8 t$. Subplot (a) shows the Fermi surface
    of $C_{{\bf k} \alpha} $ and $D_{{\bf k} \alpha} $ fermions as if there is no mixing term
    $\lambda$ in the Hamiltonian (\ref{eq:H-bs-cd}). The red line is
    $C_{{\bf k} \alpha} $ and the blue one is $D_{{\bf k} \alpha} $. Subplot (c) shows pockets
    like Fermi surface of the quasiparticles described by the
    Hamiltonian (\ref{eq:H-bs-cd}) with $\lambda = 0.3t$. Subplot (b) shows the same Fermi
    surface with the color representing the weight of electron operator in
    the quasiparticle excitation. Subplot (d) shows the same information as in (b),
    but by plotting the electron spectral
    weight at $\omega=0$ as a function of momentum; in this plot we
    manually put in a finite life-time of electron $\tau\sim0.2t$ just
    for visualization purpose. The dashed line in (a) and (c) is the
    boundary of magnetic Brillouin zone in the ordered state.}
  \label{fig:poplot}
\end{figure}
\begin{figure}[htbp]
  \centering
  \includegraphics[width=\textwidth]{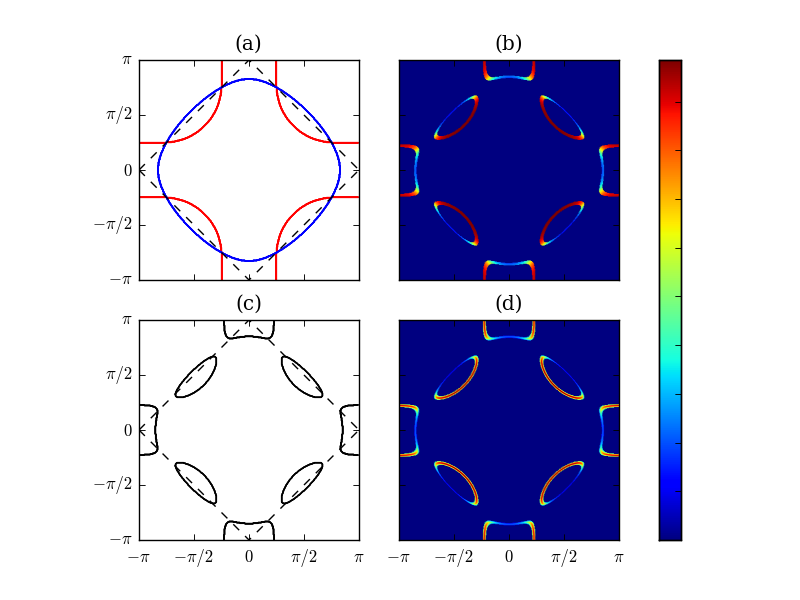}
  \caption{As in Fig.~\ref{fig:poplot} but with
    $\tilde{\varepsilon} ({\bf k}) = 0.5 t (\cos k_x+\cos k_y$) and $\lambda = 0.3t$.}
  \label{fig:poplot2}
\end{figure}
\begin{figure}[htbp]
  \centering
  \includegraphics[width=\textwidth]{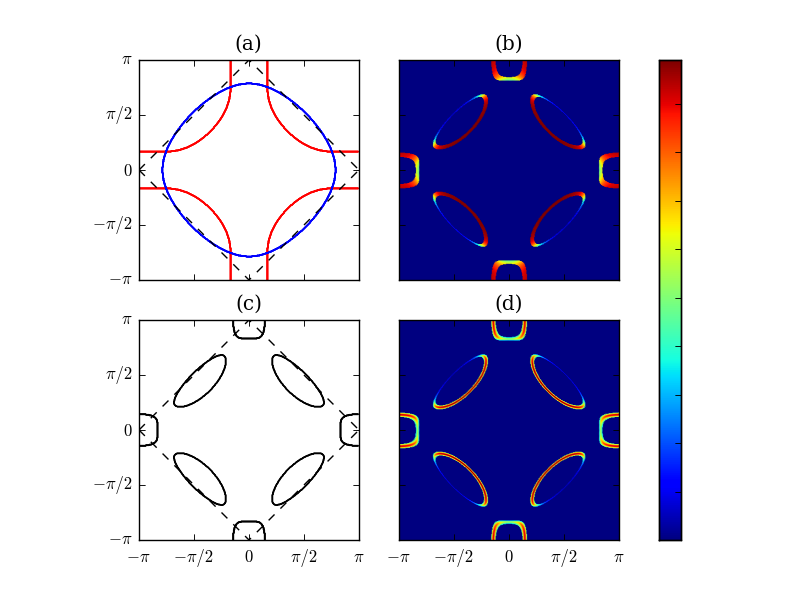}
  \caption{As in Fig.~\ref{fig:poplot} but with
    $\tilde{\varepsilon} ({\bf k}) = -0.3t + 0.5 t (\cos k_x+\cos k_y$) and $\lambda = 0.3t$.}
  \label{fig:poplot3}
\end{figure}
\begin{figure}[htbp]
  \centering
  \includegraphics[width=\textwidth]{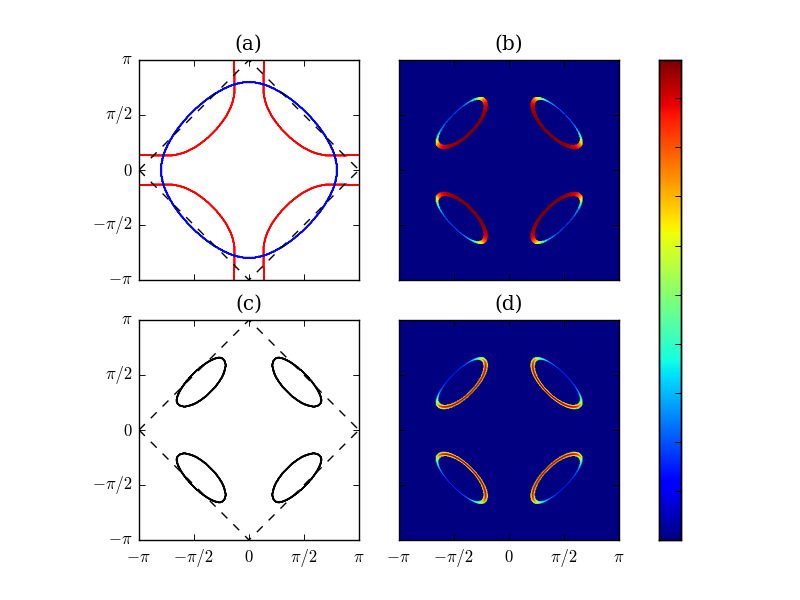}
  \caption{As in Fig.~\ref{fig:poplot} but with
    $\tilde{\varepsilon} ({\bf k}) = -0.3t + 0.5 t (\cos k_x+\cos k_y)$ and
    $\lambda=0.5t$.}
  \label{fig:poplot4}
\end{figure}

To show the qualitative effects of $\tilde{\varepsilon}$ on the shape
of Fermi pockets, we draw the pockets and electron spectrum functions
of some representative choice of $\tilde{\varepsilon} $ in
Figs. \ref{fig:poplot}-\ref{fig:poplot4}. The dispersion $\varepsilon
({\bf k})$ is chosen with some phenomenological parameters
$t^\prime=0.15t$ and $t^{\prime\prime}=-0.5t^\prime$, and the SDW gap
$\lambda$ is $0.3t$ in Figs. \ref{fig:poplot}-\ref{fig:poplot3} and
$0.5t$ in Fig. \ref{fig:poplot4}.  In Fig. \ref{fig:poplot}, a negative
$\tilde{t}_0$ shifts the hole pockets outwards, and makes the shape of
the hole pockets asymmetric. The inner side becomes more curved and
the outer side more flat. In Fig. \ref{fig:poplot2}, a negative
$\tilde{t}$ does not shift the position of the hole pockets
significantly, but also makes the shape of the hole pockets asymmetric
in a similar way as in Fig. \ref{fig:poplot}. Combining the effect of
these two parameters, we can move the hole pockets inward and make
their inner side more curved than the outer side with a position
$\tilde{t}_0$ and a negative $\tilde{t}$, as shown in
Fig. \ref{fig:poplot3}. Along the Fermi pockets, the fermionic
quasiparticles are a mixture of $C_{{\bf k} \alpha} $ and $D_{{\bf k}
  \alpha} $ fermions, while experiments only probe electron spectrum
weight. The weight of electron operator in the quasiparticle is
calculated through diagonalizing equation (\ref{eq:H-bs-cd}) and are
plotted in Fig. \ref{fig:poplot}-\ref{fig:poplot3}.  In
Fig. \ref{fig:poplot4}, we show a plot with larger $\lambda$
($\lambda=0.5t$) so that the anti-nodal electron pocket is completely
gapped. In all the cases, the inner half of the hole pockets have
higher electron quasiparticle weight, since the inner part is
primarily made of $C_{{\bf k} \alpha} $ fermion and the outer part is
primarily made of $D_{{\bf k} \alpha} $ fermion, and electron operator
is proportional to $C_{{\bf k} \alpha} $.

This bound state theory can be compared to the Yang, Rice, and Zhang (YRZ) model
of hole pockets\cite{yrz,tsvelik,lcn}. In their theory a phenomenological Green's
function for the electron in the underdoped state is proposed, based upon 
``spin liquid'' physics, and the spectral 
function derived from the Green's function has hole pockets inside the
diamond Brillouin zone. From our Hamiltonian (\ref{eq:H-bs-cd}), the
Green's function of the electron is
\begin{equation}
  \label{eq:gc}
  G^c({\bf k},\omega)=\frac{Z^2}{\omega-\varepsilon ({\bf k})-\tilde{\varepsilon} ({\bf k})+\mu
    +\lambda^2/(\omega-\varepsilon ({\bf k} + {\bf K}) +\tilde\varepsilon ({\bf k} + {\bf K}) +\mu)}
\end{equation}
Compare this to the Green's function in the YRZ model \cite{yrz,tsvelik,lcn}:
\begin{equation}
  \label{eq:yrz}
  G({\bf k},\omega)=\frac{g_t}{\omega-\xi(k)-\Delta_k^2/[\omega+\xi_0(k)]}.
\end{equation}
The two results are quite similar. In the present form, instead of the
$d$-wave gap $\Delta_k$, our
$\lambda$ does not have a momentum dependence. However, this distinction is an artifact of the simple 
choices made in our form of $H_{\rm eff}$: we can clearly include more non-local mixing between the
$F$ and $G$ fermions.

The more important distinction between our model and YRZ lies in the physical input
in the fermion spectrum. The YRZ model relies on `pairing correlations' implicit in some underlying
spin liquid state. In contrast, we do not assume any pairing, but fluctuating local antiferromagnetic
order. We will describe the influence of pairing on our spectral functions in a future paper.
We note earlier arguments \cite{gs} that the pairing amplitudes are especially strong on the electron pockets,
and this may be a contributing factor to removing the electron-like Fermi surfaces in Figs.~\ref{fig:poplot2}
and~\ref{fig:poplot3}.

Recent ARPES experiments \cite{pocket3} reveal that there are hole pockets in the
underdoped regime, and the center of the pockets are inside the first
magnetic Brillouin zone. In general the shape of hole pockets seen in
the experiments can be fit to our model. As shown in
Fig. \ref{fig:poplot}-\ref{fig:poplot4}, the weight of electron
operator on the outer part of the pockets is tiny and may be hard to
see in experiments.

We close this section by remarking on the status of Luttinger's
theorem in our theory of the electron spectral function. These issues
were discussed in Refs.~\onlinecite{rkk2,rkk3}, where we argued that
the theorem applied to the sum of the $F_\alpha$, $G_\alpha$ and the
$\psi_{\pm}$ Fermi surfaces.  Here we will focus on the electron-like
$F_\alpha$, $G_\alpha$ , and will drop the $\psi_\pm$ contributions to
the present discussion; the latter will amount to shift in the
effective doping level $x$. Under this assumption the total number of
$F_\alpha$ and $G_\alpha$ per site is $2-x$, and so is the number of
$C_\alpha$ and $D_\alpha$, since the canonical transformation in
equation (\ref{eq:CD}) preserves particle number. On the other hand,
our theory was applied to the doped holes or electrons in the
background of a fluctuating antiferromagnet, and each such charge
carrier must occupy one state within the Fermi surface.  Counting hole
(electron)-like Fermi surfaces as negative (positive), then for a
doped antiferromagnet with hole density $x$, the Fermi surfaces in
Figs.~\ref{fig:poplot}-\ref{fig:poplot4} should enclose a total area
of $- (2 \pi)^2 (x/2)$ (the last factor of 2 is from spin
degeneracy). In our present effective Hamiltonian model, we have found
it more efficient to treat not just the fermionic excitations near the
electron and hole pockets, but across the entire Brillouin zone. We
found that this method was very convenient in treating the contraints
imposed by symmetry, without prejudicing the final locations of the
Fermi surfaces. Such an extension should not be accompanied by any
fundamental change in the many body quantum state, and hence cannot
modify the statement of Luttinger's theorem.  Because the Luttinger
constraint only controls the electron density modulo 2 per unit cell,
we therefore conclude that our lattice effective model obeys
\begin{equation}
\left\langle F_{i \alpha}^\dagger F_{i \alpha} \right\rangle + \left\langle G_{i \alpha}^\dagger G_{i \alpha} \right\rangle
= \left\langle C_{i \alpha}^\dagger C_{i \alpha} \right\rangle + \left\langle D_{i \alpha}^\dagger D_{i \alpha} \right\rangle
= 2 -x \label{luttth}
\end{equation}
Note that this differs from the value in the conventional Fermi liquid phase, in which case the total electron density is $1-x$.
This difference is acceptable here because we are discussing a phase with topological order, which has an emergent
U(1) gauge excitation \cite{ffl}.

In the discussion of YRZ model\cite{yrz,tsvelik} a different form of
Luttinger's theorem is used: there the total number of particles given
by Green's function (\ref{eq:yrz}) is given by the total area enclosed
by the contours where the Green's function changes sign. This includes
not only the Fermi pockets on which the Green's function diverges, but
also the contour where the Green's function vanishes. For our Green's
function of $C$ bound states in equation (\ref{eq:gc}) this extra
Luttinger surface of zeros is exactly the original Fermi surface of
$D$ bound states without the mixing between $C$ and $D$. Therefore the
total area enclosed by both Fermi pockets and surface of zeros equal
to total number of $C$ and $D$ states minus the number of $D$ states,
and the result is the total number of $C$ states. So this form of
Luttinger's theorem still holds true in our model, although it is not
relevant to the physical doping, which is related to the area of Fermi
pockets only.

\subsection{Coupling to photons}
\label{sec:photons}

Apart from the electron-like bound states described in the previous section, the other low-lying excitations
in the ACL are the gapless U(1) photons. Here we discuss how these two low-lying sectors of the theory couple to each other.

Our first task is search for terms coupling the $F_\alpha$, $G_\alpha$ to the photons, while being invariant under all the
transformations in Table~\ref{table0}. It is useful to do this on the lattice, as many of the operations involve details of
the lattice symmetry. The $F_\alpha$, $G_\alpha$ are gauge-invariant, and so will couple only to the field strengths: on the lattice,
we define these as
\begin{equation}
\mathcal{B} = \Delta_x A_y - \Delta_y A_x~~,~~\mathcal{E}_x = \Delta_x A_\tau - \Delta_\tau A_x~~,~~\mathcal{E}_y = \Delta_y A_\tau - \Delta_\tau A_y;
\end{equation}
thus $\mathcal{B}$ resides at the center of each square lattice plaquette, while $\mathcal{E}_{x,y}$ reside on the links.

After some searching, we found a single term which is linear in the field strengths which fulfills the needed criteria:
\begin{equation}
\label{eq:gamma}
\mathcal{S}_\gamma = - i \gamma \int d \tau \sum_i  \left[ \mathcal{E}_x \left( F_\alpha^\dagger \Delta_x G_\alpha - G_\alpha^\dagger \Delta_x F_\alpha \right) + \mathcal{E}_y \left( F_\alpha^\dagger \Delta_y G_\alpha - G_\alpha^\dagger \Delta_y F_\alpha \right) \right] 
\end{equation}

\begin{figure}[htbp]
  \centering
  \includegraphics{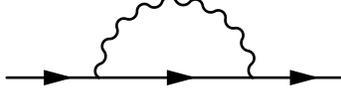}
  \caption{Feynman diagram for the self-energy correction. The solid
    line with arrow represents $F$ or $G$ fermion propagator and the
    wiggly line represents electric field propagator. The interaction
    vertex is given by equation (\ref{eq:gamma}). This diagram is
    evaluated in equation (\ref{eq:self-energy}).}
  \label{fig:self-energy}
\end{figure}

This coupling contributes to the self-energy correction of the bound
states. At the lowest order, the self-energy correction is represented
by the Feynman diagram in Fig. \ref{fig:self-energy},

The action of U(1) gauge field has contributions from both the
$\psi_\pm$ fermions and $z_\alpha$ field. Assuming that there exist
Fermi surfaces of the $\psi_\pm$ fermions, integrating them out screens
the fluctuation of the $A_\tau$ component, and gives
the following terms for the action of the transverse components of $A$
\begin{equation}
  \label{eq:SAf}
  S^{Af}=\frac{1}{2} \int\frac{d^2pd\omega}{(2\pi)^3}
  A_i(-p,-\omega)\left(\delta^{ij}-\frac{p^ip^j}{p^2}\right)
  \left(k_F^\psi\frac{|\omega|}{p}+p^2 \right)A_j(p,\omega)
\end{equation}
Here $k_F^\psi$ is the Fermi momentum of the
$\psi_\pm$ Fermi surface.

We consider two possibilities of the $z_\alpha$ fields. At the magnetic
ordering critical point, integrating out $z_\alpha$ field gives rise
to the following action
\begin{equation}
  \label{eq:SAv}
  S^{Az}=\frac{1}{2} \int\frac{d^2pd\omega}{(2\pi)^3}
  A_i(-p,-\omega)\left(\delta^{ij}-\frac{p^ip^j}{p^2}\right)
  \sqrt{\omega^2+c_s^2p^2}A_j(p,\omega)
\end{equation}
On the other hand, if the $z_\alpha$ field is gapped as in the ACL,
integrating it out gives rise to terms proportional to
$\omega^2+c_s^2p^2$, which are higher order  than those in
Eq~(\ref{eq:SAf}).

In summary, the U(1) gauge field action can be written as
\begin{equation}
  \label{eq:SA}
  S^{Az}=\frac{1}{2} \int\frac{d^2pd\omega}{(2\pi)^3}
  A_i(-p,-\omega)\Pi(p, \omega)
  A_j(p,\omega)
\end{equation}
where the polarizition function $\Pi$ is given by
\begin{align}
  \label{eq:Pi-crt}
  &\Pi(p,
  \omega)=k_F^\psi\frac{|\omega|}{p}+\sqrt{\omega^2+c_s^2p^2}
  +\cdots
  &\text{at the quantum critical point}\\
  \label{eq:Pi-dis}
  &\Pi(p,
  \omega)=k_F^\psi\frac{|\omega|}{p}
  + p^2
  &\text{in the ACL}
\end{align}
for imaginary frequencies $\omega$.

Using the action in equation~\eqref{eq:SAf} and \eqref{eq:SAv} we can
evaluate the Feynman diagram in Fig.~\ref{fig:self-energy}.
\begin{equation}
  \label{eq:self-energy}
  \Sigma(p,\omega)=\gamma^2\int\frac{d^2qd\Omega}{(2\pi)^3}
  \frac{1}{i(\Omega+\omega)-\xi_q}
  \frac{\Omega^2}{\Pi(|\bm{p}-\bm{q}|,\Omega)}
  (p_i+q_i)(p_j+q_j)\left[\delta^{ij}-\frac{(p_i-q_i)(p_j-q_j)}{(\bm{p}-\bm{q})^2}
  \right]
\end{equation}
Here we consider a simple one-band model for the bound state with
a quadratic dispersion $\xi_a={q^2}/({2m})-\mu$. The polarization
factor at the end can be simplified as
\begin{equation}
  \label{eq:polar}
  (p_i+q_i)(p_j+q_j)\left[\delta^{ij}-\frac{(p_i-q_i)(p_j-q_j)}{(\bm{p}-\bm{q})^2}
  \right]
  =\frac{4p^2q^2(1-\cos^2\theta)}{p^2+q^2-2pq\cos\theta}
\end{equation}
where $\theta$ is the angle between $\bm{p}$ and $\bm{q}$.

We are interested in the life-time of quasiparticles near the Fermi
surface, so we consider the imaginary part of self-energy at the
quasiparticle pole $(p,\omega=\xi_p)$. Without losing generality, we
consider the case of $\omega>0$. At zero temperature, energy of the
intermediate fermion state must be greater than zero but smaller than
$\omega$, so its momentum must be inside the shell of $k_F<q<p$. The
imaginary part of self-energy is
\begin{equation}
  \label{eq:self-energy-img}
  \begin{split}
  \text{Im}\Sigma^{\text{ret}}(p, \omega)=&\frac{\gamma^2}{(2\pi)^2}\int_{k_F}^pqdq
  \int_0^{2\pi} d\theta (\omega-\xi_q)^2\\
  &\text{Im}\frac{1}{\Pi^{\text{ret}}(|\bm{p}-\bm{q}|, \omega-\xi_q)}
  \frac{4p^2q^2(1-\cos^2\theta)}{p^2+q^2-2pq\cos\theta}
  \end{split}
\end{equation}

Near the Fermi surface we can linearize the dispersion relation
$\xi_q=v_F(q-p_F)$, and $\omega=\xi_p=v_F(p-p_F)$. We change the
integrated variable from $q$ to $k=p-q$, $0<k<\omega/v_F$.  In the
limit of $\omega\rightarrow0$, or $p-p_F\rightarrow0$, the integral
can be simplified 
\begin{equation}
  \label{eq:self-energy-img-2}
  \text{Im}\Sigma^{\text{ret}}(p,
  \omega)=\frac{\gamma^2k_F^6}{\pi^2m^2}
  \int_0^{\frac{\omega}{v_F}}dk
  \int_0^{2\pi} d\theta k^2
  \text{Im}\frac{1}{\Pi^{\text{ret}}(
    \sqrt{k^2+2p_F^2(1-\cos\theta)}, v_Fk)}
  \frac{1+\cos\theta}{k^2/(1-\cos\theta)+p_F^2}
\end{equation}
In both of the cases considered in Eqs.~\eqref{eq:Pi-crt}
and~\eqref{eq:Pi-dis}, the gauge field spectrum function
\[\text{Im}\frac{1}{\Pi^{\text{ret}}(
  \sqrt{k^2+2p_F^2(1-\cos\theta)}, v_Fk)}\sim k^{-1},\quad
k\rightarrow0\] when $\theta\neq0$, and becomes less singular at
$\theta=0$. As a result the integral in Eq.~\eqref{eq:self-energy-img} has the following behavior
\[\text{Im}\Sigma^{\text{ret}}(p,\omega)\sim\omega^2,\quad
\omega\rightarrow0 . \] Thus the bound state fermion has a Fermi
liquid-like damping.

This conclusion is different from Essler and Tsvelik's work on a
model of weakly coupled chains \cite{essler02, essler05}. Their pocket states
arose from binding between the one-dimensional holons and spinons, and the
coupling to the one-dimensional spin fluctuations led to a
self-energy proportional to $\omega \ln(\omega)$. Our model is genuinely two-dimensional,
and has an emergent U(1) gauge field who presence is also the key to the 
violation of the Luttinger theorem in Eq.~(\ref{luttth}).
They explain the violation of the Luttinger theorem by zeros of the Green's function \cite{essler02,essler05,tsvelik};
the connection of these zeros to our work was discussed below Eq.~(\ref{luttth}).

\section{Electron spectral functions in the renormalized classical regime}
\label{sec:rc}

We will now consider the regime at small non-zero $T$ above the antiferromagnetically ordered state
present at $g<g_c$, the renormalized classical (RC) regime.
Here, we expect the O(3) vector formulation of $\mathcal{L}_{sf}$ in Eq.~(\ref{esf}) 
to be equivalent to the spinor/U(1) gauge theory formulation of $\mathcal{L}_{acl}$ in Eq.~(\ref{lacl}). 
This is explored at $T=0$ in Appendix~\ref{smallg}, where we show that the small $g$ expansions
of the two theories match with each other. Thus, while there are differences in the nature of the
non-magnetic phases at $g>g_c$ in the two cases (as reviewed in Section~\ref{sec:intro}), the $g<g_c$
phase and its low $T$ RC regimes are the same.

The RC regime was studied using the O(3) formulation by Vilk and Tremblay \cite{vilk},
and using a U(1) gauge theory similar to $\mathcal{L}_{acl}$ by Borejsza and Dupuis \cite{borejsza}.
Here we shall also use $\mathcal{L}_{acl}$, and expand upon these earlier results.

We compute the electron spectral function as a convolution
of the $z_\alpha$ propagator with the free $\psi$ propagator containing pocket Fermi surfaces obtained from
$\mathcal{L}_\psi$.

For the $z_\alpha$ propogator we use the simple damped form motivated by studies in the $1/N$ expansion.\cite{sscsy}
\begin{equation}
G_z ({\bf k}, \omega) = \frac{1}{-(\omega + i \gamma m)^2 + v^2 k^2  + m^2} \label{Gz}
\end{equation}
where the dimensionless constant $\gamma$ determines the damping constant, and the ``mass'' $m$ is determined by
the ``large $N$'' equation
\begin{equation}
m = 2 T \ln \left[ \frac{ e^{- 2 \pi \varrho /T} + \sqrt{ e^{- 4 \pi \varrho/T} + 4}}{2} \right]
\end{equation}
where
\begin{equation}
\varrho = v^2 \left( \frac{1}{g} - \frac{1}{g_c} \right)
\end{equation}
is the energy scale which determines the deviation from the quantum
critical point.  The damping in Eq.~(\ref{Gz}) is a simple
interpolation form which is constant with the expected behavior at the
quantum critical point, and with the RC behavior of Eq.~(7.28) in Ref.~\cite{ssbook}. 
Actually, we have neglected the power-law pre-factor of the
exponential in (7.28).  Determining the damping from numerical studies
motivated by the $\epsilon$ expansion \cite{sscrit} we expect $\gamma
\approx 1$.

\begin{figure}[htbp]
  \centering
  \includegraphics[width=5in]{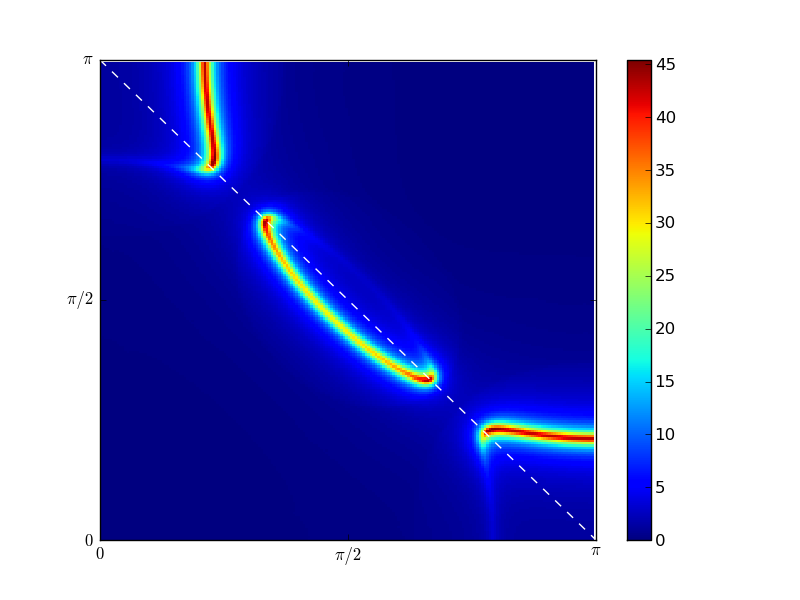}
  \caption{Electron spectrum function at zero frequency as a
    convolution of $z_\alpha$ and the $\psi_{\pm}$ fermions. In the plot we chose the following
    parameters: $\varrho=0.1t$, $T=0.2t$ and the SDW gap $\lambda=0.2t$. The white dashed line shows the
    boundary of the diamond magnetic Brillouin zone of the
    commensurate $(\pi,\pi)$ AFM order.}
  \label{fig:conv}
\end{figure}

\begin{figure}[htbp]
  \centering
  \includegraphics[width=5in]{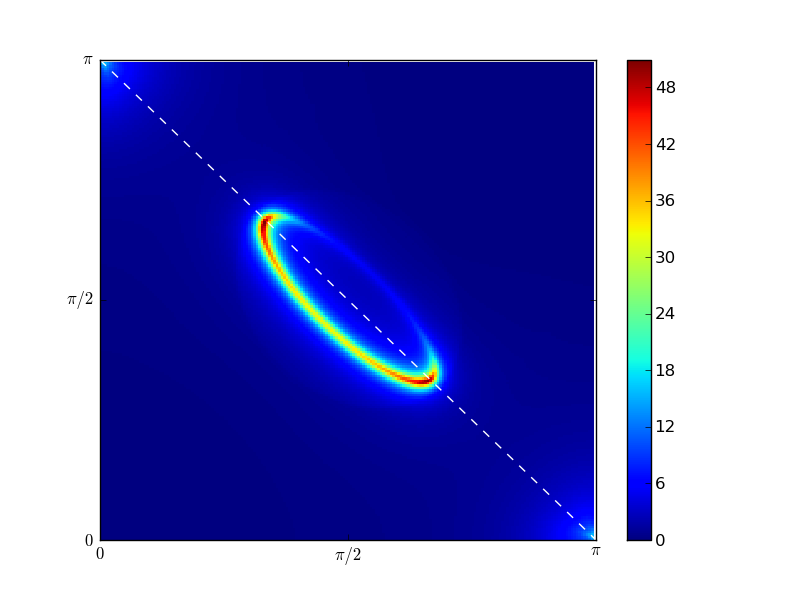}
  \caption{Electron spectrum function at zero frequency as a
    convolution of $z_\alpha$ and the $\psi_{\pm}$ fermions. In the
    plot we chose the following parameters: $\varrho=0.1t$, $T=0.2t$
    and the SDW gap $\lambda=0.5t$. The white dashed line shows the
    boundary of the diamond magnetic Brillouin zone of the
    commensurate $(\pi,\pi)$ AFM order.}
  \label{fig:conv2}
\end{figure}

In Fig.~\ref{fig:conv} and \ref{fig:conv2} we plot electron spectrum
function obtained as a convolution of the $z_\alpha$ and fermion
spectrum functions at finite temperature. Here fermion dispersion
relation is the same as the one we used in the plots in section
\ref{sec:bound}. In contrast to
Fig. \ref{fig:poplot}-\ref{fig:poplot4}, the pockets are symmetric
with respect to the magnetic Brillouin zone boundary, and hole pockets
are centered at $(\pi/2,\pi/2)$. Again, the inner half of the pockets
has higher spectrum weight, as the quasiparticle on outer half of the
pockets is primarily made of fermion quasiparticles at
$f_{k+(\pi,\pi)}$. The broadening of the spectrum comes from two
factors: first, the $z$ boson has a finite life-time $\gamma m$;
second, the convolution is done at finite temperature, so the energy
of fermion and boson excitation does not have to be exactly zero, but
can vary by energies of order $T$. The second factor dominantes in the
RC regime because $m\ll T$, and $\gamma$ is of order 1. So the
electron spectrum is basically the pocket fermion dispersion broadened
by a linewidth of order $T$: see Fig.~\ref{fig:conv} and
Fig.~\ref{fig:conv2}.
Earlier work \cite{vilk} had also obtained a linewidth of order $T$ 
in the RC regime by very different methods.

As we noted above, the present Fermi surface locations are symmetric with respect to the 
magnetic Brillouin zone boundary, unlike those in 
Figs. \ref{fig:poplot}-\ref{fig:poplot4}. However, as shown in Ref.~\onlinecite{rkk1}, the 
Shraiman-Siggia term in Eq.~(\ref{lss} lifts this symmetry. We can compute perturbative corrections
to leading order in $\lambda$, and these will contribute spectral weight of width $T$
which is asymmetric about the magnetic Brillouin zone boundary.

\section{Conclusions}
\label{sec:conc}

We conclude by reviewing the different routes \cite{su2} by which the spin fermion model $\mathcal{L}_{sf}$
in Eq.~(\ref{esf}) can lose antiferromagnetic order, and their implications for the photoemission spectrum
at finite temperatures.

The first, more conventional, route is that there is a single direct transition at $g=g_c$ to a Fermi liquid
with a large Fermi surface. Then in the $T$-$g$ plane, we have the conventional \cite{ssbook}
quantum disordered (QD), quantum critical (QC), and renormalized classical (RC) regions.
In QD region at small $T$ for $g > g_c$, the electron spectrum will show quasiparticle peaks with 
a Fermi liquid linewidth $\sim T^2$. In the QC region near $g = g_c$, the spectrum will again show
weight along the large Fermi surface, but with large anomalous linewidths near the `hot spots':
these are points along the large Fermi surface connected by the ordering wavevector ${\bf K}$.
Finally, in the RC region at small $T$ for $g<g_c$, we have the behavior described in
Section~\ref{sec:rc}: the spectrum has `small' Fermi pockets which are centered at the antiferromagnetic
Brillouin zone boundary, and the quasiparticle peaks have a width $\sim T$; examples of such
spectra were shown in Figs.~\ref{fig:conv}, \ref{fig:conv2}.

This paper mainly considered a more exotic route \cite{rkk1,rkk2} towards loss of antiferromagnetic
order in the spin-fermion model. This route is possible if topological ``hedgehog'' tunnelling events
are suppressed as the transition, as they are at magnetic-disordering transitions in the insulator
(see Fig.~\ref{figrkk}). Then the transition at $g=g_c$ is to a non-magnetic non-Fermi liquid
ACL, with the spinless fermions $\psi_\pm$ and complex bosonic spinons $z_\alpha$ as elementary excitations,
interacting (in the simplest case) via an emergent U(1) gauge force. Again, around this critical point
at $g=g_c$, we can define the corresponding RC, QC, and QD regions. The electron spectrum in the RC
region is just as in the first case above, as discussed in Section~\ref{sec:rc} and Appendix~\ref{smallg}.
The novel spectrum in the QD region was the main focus of Section~\ref{sec:bound}.
One contribution to the spectral weight comes from the convolution of the deconfined spinons
and the $\psi_{\pm}$. This leads to incoherent spectral weight in `arc'-like regions which were described
earlier in Ref.~\onlinecite{rkk1}. However, it has been argued \cite{rkk1,rkk2} that the spinons and $\psi_\pm$ form
electron-like bound states, and these were described in more detail in Section~\ref{sec:bound}. We found
that the bound states lead to pocket Fermi surfaces, as shown in Figs~\ref{fig:poplot}-\ref{fig:poplot4}.
An important feature of these spectra are that the pockets are not centered at the point $(\pi/2,\pi/2)$ on the
antiferromagnetic Brillouin zone boundary. In fact, this zone boundary plays no special role, and there
are no symmetry relations on the quasiparticle dispersions across it. We note that recent
photoemission observations \cite{pocket3} show features related to this QD region. We did not address the QC
region here, but it should mainly have the incoherent arc spectra, as described in Ref.~\onlinecite{rkk1}.

We also noted a similarity between our ACL QD results, and the YRZ model \cite{yrz,tsvelik,lcn} in Section~\ref{sec:bound}. The assumptions of YRZ are very different from ours, as they depart from a  paired fermionic spinons
in a spin-liquid state. The Fermi surfaces in the YRZ model to not obey the conventional Luttinger theorem, as is the case
in our model. However, such a violation must be accompanied by gauge forces reflecting the topological order in such a state \cite{ffl}:
these gauge fields do not appear in their formulation.
Also, we dealt mainly with the influence of local antiferromagnetic order on the electron spectrum.
We have not included pairing effects in our computations yet, or the transition to superconductivity; these are issues
we hope to address in forthcoming work.

\acknowledgments

We thank J.~Carbotte, M.~Metlitski, and A.-M.~Tremblay for useful discussions.
This research was supported by the National Science Foundation under grant DMR-0757145, by the FQXi
foundation, and by a MURI grant from AFOSR.

\appendix

\section{Small $g$ expansion}
\label{smallg}

This Appendix will compare the properties of the Lagrangian $\mathcal{L}_{sf}$ defined in Eq.~(\ref{esf}),
with the `fractionalized' Lagrangian $\mathcal{L}_{acl}$ in Eq.~(\ref{lacl}). We will work at $T=0$ in the
limit of small $g$, where both models have long range SDW order, and are expected to be essentially
identical. We will compare the two models here by computing the on-shell matrix element for the decay of 
a spin-wave in the SDW state into a fermionic particle hole pair.

We will work in the ``diamond'' Brillouin zone associated with antiferromagnetic ordering with
wavevector ${\bf K} = (\pi, \pi)$. In this zone we define
\begin{eqnarray}
c_{1 \alpha} ({\bf k}) \equiv c_{\alpha } ({\bf k})~~~&,&~~~\varepsilon_1 ({\bf k}) = \varepsilon ({\bf k}) \nonumber \\
c_{2 \alpha} ({\bf k}) \equiv c_{\alpha } ({\bf k} + {\bf K})~~~&,&~~~\varepsilon_2 ({\bf k}) = \varepsilon ({\bf k} + {\bf K}) 
\end{eqnarray}
All expressions below are implicitly in this diamond Brillouin zone.

The analysis of the spin-wave decay appears for the two models in the following subsections.

\subsection{Spin-fermion model}
\label{app:sf}

We perform the small $g$ expansion for the order parameter by the following parameterization in terms of the spin-wave
field $\phi$:
\begin{equation}
n^a = \left(  \frac{\phi + \phi^\ast}{2} \sqrt{2g - g^2 |\phi|^2},  \frac{\phi - \phi^\ast}{2i}\sqrt{2g - g^2 |\phi|^2}, 1 - g |\phi|^2 \right)
\end{equation}
Then the Lagrangian for $n^a$ is
\begin{equation}
\mathcal{L}_n =  |\partial_\mu \phi |^2 + \frac{g}{4}\left[ \left( \phi^\ast \partial_\mu \phi \right)^2 + \left( \phi \, \partial_\mu \phi^\ast \right)^2 \right]
+\frac{g^2}{8} \frac{|\phi|^2}{(2- g |\phi|^2)} \left( \phi^\ast \partial_\mu \phi + \phi \, \partial_\mu \phi^\ast \right)^2 \label{lnp}
\end{equation}
and can be analyzed as usual in an expansion in $g$. For the fermion sector, we diagonalize the $\mathcal{L}_c + \mathcal{L}_\lambda $ at $g=0$
by introducing fermion operators $\gamma_{1,2p}$, and (we replace the electron index $\alpha$ by $p=\pm 1$)
\begin{eqnarray}
c_{1p} ({\bf k}) &=& u_{\bf k} \gamma_{1 p} ({\bf k}) - p v_{\bf k} \gamma_{2 p} ({\bf k})  \nonumber \\
c_{2p} ({\bf k})  &=& p v_{\bf k} \gamma_{1 p} ({\bf k}) + u_{\bf k} \gamma_{2 p} ({\bf k})
\end{eqnarray}
where $u_{\bf k}$, $v_{\bf k}$ are real and obey $u_{\bf k}^2 + v_{\bf k}^2=1$. We choose $u_{\bf k} = \cos (\theta_{\bf k}/2)$, $v_{\bf k} 
= \sin (\theta_{\bf k}/2)$, and then
\begin{equation}
\cos \theta_{\bf k} = \frac{\varepsilon_1 ({\bf k}) - \varepsilon_2 ({\bf k})}{\sqrt{ (\varepsilon_1 ({\bf k}) - \varepsilon_2 ({\bf k}))^2 + 4 \lambda^2}}
~~~\sin \theta_{\bf k} = \frac{-2 \lambda}{\sqrt{ (\varepsilon_1 ({\bf k}) - \varepsilon_2 ({\bf k}))^2 + 4 \lambda^2}}.
\end{equation}
The fermion Lagriangian at $g=0$ is 
\begin{equation}
\mathcal{L}_\gamma = \sum_{{\bf k}, p} \gamma^{\dagger}_{1 p} \left( \partial_\tau + E_1 ({\bf k}) \right) \gamma_{1 p}
+ \sum_{{\bf k}, p} \gamma^{\dagger}_{2 p} \left( \partial_\tau + E_2 ({\bf k}) \right) \gamma_{2 p}
\end{equation}
where
\begin{equation}
E_{1,2} ({\bf k}) = \frac{1}{2} \left[ \varepsilon_1 ({\bf k}) + \varepsilon_2 ({\bf k}) \pm \sqrt{ (\varepsilon_1 ({\bf k}) - \varepsilon_2 ({\bf k}))^2 + 4 \lambda^2} \right]
\end{equation}
Finally, the non-linear couplings between the spin waves and the fermions is given by
\begin{eqnarray}
\mathcal{L}_{\phi,\gamma} &=& - \lambda \sqrt{2g} \sum_{{\bf k}, {\bf q}} 
\left[ \phi \sqrt{1 - g |\phi|^2/2} \right]_{\bf q} \Bigl[ \nonumber \\
&~&~~~~~~~~~~~~~~~~(u_{{\bf k}+{\bf q}} v_{{\bf k}} - u_{\bf k} v_{{\bf k} + {\bf q}} ) (
\gamma_{1-}^\dagger ({\bf k} + {\bf q}) \gamma_{1+} ({\bf k}) - \gamma_{2-}^\dagger ({\bf k} + {\bf q}) \gamma_{2+} ({\bf k})) \nonumber \\
&~&~~~~~~~~~~~~+  (u_{{\bf k}+{\bf q}} u_{{\bf k}} + v_{\bf k} v_{{\bf k} + {\bf q}} ) (
\gamma_{1-}^\dagger ({\bf k} + {\bf q}) \gamma_{2+} ({\bf k}) + \gamma_{2-}^\dagger ({\bf k} + {\bf q}) \gamma_{1+} ({\bf k})) \Bigr] + \mbox{H.c.}
\nonumber \\
&~& + g \lambda \sum_{{\bf k}, {\bf q}} \sum_p
\left[ |\phi|^2 \right]_{\bf q} \Bigl[ (u_{{\bf k}+{\bf q}} v_{{\bf k}} + u_{\bf k} v_{{\bf k} + {\bf q}} ) (
\gamma_{1p}^\dagger ({\bf k} + {\bf q}) \gamma_{1p} ({\bf k}) - \gamma_{2p}^\dagger ({\bf k} + {\bf q}) \gamma_{2p} ({\bf k})) \nonumber \\
&~&~~~~~~~~~~~~+  p (u_{{\bf k}+{\bf q}} u_{{\bf k}} - v_{\bf k} v_{{\bf k} + {\bf q}} ) (
\gamma_{1p}^\dagger ({\bf k} + {\bf q}) \gamma_{2p} ({\bf k}) + \gamma_{2p}^\dagger ({\bf k} + {\bf q}) \gamma_{1p} ({\bf k})) \Bigr]
\end{eqnarray}

Now we can obtain the self-energy of the $\phi$ spin-wave excitation to order $g$:
\begin{eqnarray}
&& \Sigma_{\phi} ({\bf q}, \omega_n) = - 2 \lambda^2 g \sum_{{\bf k}} \Biggl[ \nonumber \\
&&~~~~~~(u_{{\bf k}+{\bf q}} v_{{\bf k}} - u_{\bf k} v_{{\bf k} + {\bf q}} )^2 \left( \frac{ f(E_1 ({\bf k})) - f (E_1 ({\bf k}+{\bf q}))}{- i \omega_n  + E_1 ({\bf k} + {\bf q}) - E_1 ({\bf k})} +  \frac{ f(E_2 ({\bf k})) - f (E_2 ({\bf k}+{\bf q}))}{- i \omega_n  + E_2 ({\bf k} + {\bf q}) - E_2 ({\bf k})} \right) \nonumber \\
&&~+~(u_{{\bf k}+{\bf q}} u_{{\bf k}} + v_{\bf k} v_{{\bf k} + {\bf q}} )^2 \left( \frac{ f(E_2 ({\bf k})) - f (E_1 ({\bf k}+{\bf q}))}{- i \omega_n  + E_1 ({\bf k} + {\bf q}) - E_2 ({\bf k})} +  \frac{ f(E_1 ({\bf k})) - f (E_2 ({\bf k}+{\bf q}))}{- i \omega_n  + E_2 ({\bf k} + {\bf q}) - E_1 ({\bf k})} \right) \Biggr]
\nonumber \\
&&~~~~~~~~~~~~~+ 2 g \lambda \sum_{{\bf k}} 2 u_{\bf k} v_{\bf k} ( f(E_1({\bf k})) - f(E_2 ({\bf k}))) 
\end{eqnarray}
As expected, we have $\Sigma_\phi (0,0) = 0$. We can also estimate the on-shell decay rate at $T=0$:
$\mbox{Im} \Sigma_\phi (vq, q) \sim g q^2$, as long as $v < v_F$.

We can also use $\mathcal{L}_{\phi,\gamma}$ to compute the fermion spectral density at order $g$. For the case
of a single particle in the insulator, it seems to me that the result has the same form as (2.47) in Ref.~\onlinecite{klr}.

\subsection{U(1) gauge theory}

For small $g$, 
we parametrize the $z_\alpha$ field of the CP$^1$ model as
\begin{equation}
z_\alpha = \left( \begin{array}{c} \sqrt{1 - g |\phi|^2/2} \\ \sqrt{g/2} \, \phi \end{array} \right) e^{i \vartheta} \label{zp}
\end{equation}
Then the Lagrangian for $z_\alpha$ is
\begin{equation}
\mathcal{L}_z =  |\partial_\mu \phi |^2  
+\frac{g}{4(2- g |\phi|^2)} \left( \phi^\ast \partial_\mu \phi + \phi \, \partial_\mu \phi^\ast \right)^2
+ \frac{2}{g} ( \partial_\mu \vartheta - A_\mu)^2 -i ( \partial_\mu \vartheta - A_\mu) ( \phi^\ast \partial_\mu \phi - \phi \, \partial_\mu \phi^\ast )
\end{equation}
If we integrate over $A_\mu$ in the gauge $\vartheta=0$, we reproduce the action in Eq.~(\ref{lnp}).

The integral over $A_\mu$ also produces a coupling between $\phi^\ast \partial_\mu \phi - \phi \, \partial_\mu \phi^\ast$
and the U(1) current over the $\psi_{1,2p}$ fermions. Note that this is a quartic coupling, a bilinear in $\phi$ coupling
to a bilinear in $\psi_{1,2p}$. 

In our discussion in Section~\ref{app:sf}, the dominant coupling between the fermions and the spin-waves was the
term in $\mathcal{L}_{\phi,\gamma}$ which was {\em linear\/} in $\phi$. Such terms arise here
exclusively from $\mathcal{L}_{ss}$. We evaluated the on-shell matrix elements from such terms for the decay of a $\phi$ spin wave into 
a particle hole pair of the $\gamma_{1p}$ (and also to the $\gamma_{2p}$). From the first term in $\mathcal{L}_{ss}$ we
have the matrix element
\begin{equation}
\sqrt{g/2}\, {\bf q} \cdot \left( u_{\bf k}^2 \frac{ \partial \varepsilon_1 ({\bf k})}{\partial {\bf k}} -  
v_{\bf k}^2 \frac{ \partial \varepsilon_2 ({\bf k})}{\partial {\bf k}} \right)
\label{m1}
\end{equation}
For the time-derivative term in $\mathcal{L}_{ss}$ we replace the frequency by its on-shell value ${\bf q}\cdot \partial E_1 ({\bf k})/\partial {\bf k}$ to obtain the matrix element
\begin{equation}
-\sqrt{g/2} \, {\bf q} \cdot \frac{ \partial E_1 ({\bf k})}{\partial {\bf k}}  \left( u_{\bf k}^2  -  
v_{\bf k}^2 \right) 
\label{m2}
\end{equation}
The sum of (\ref{m1}) and (\ref{m2}) is equal to the matrix element obtained in Section~\ref{app:sf}, which is
\begin{equation}
\lambda \sqrt{2g} \, {\bf q} \cdot \left( v_{\bf k} \frac{\partial u_{\bf k}}{\partial {\bf k}} - u_{\bf k} \frac{\partial v_{\bf k}}{\partial {\bf k}}
\right).
\end{equation}

\end{document}